\documentclass{article}

\usepackage[hang,flushmargin]{footmisc}
\usepackage{hyperref}

\usepackage[accepted]{icml2026}

\usepackage[utf8]{inputenc}
\usepackage[T1]{fontenc}
\usepackage{amsmath,amssymb,amsthm}
\usepackage{mathtools}
\usepackage{booktabs}
\usepackage{tikz-cd}
\usepackage{float}
\usepackage{placeins}
\usepackage{color}
\usepackage{xcolor}
\usepackage{listings}
\usepackage{textcomp}
\usepackage{fontawesome5}

\definecolor{lightblue}{rgb}{0.8,0.85,1}
\definecolor{keywordcolor}{rgb}{0.7, 0.1, 0.1}
\definecolor{tacticcolor}{rgb}{0.0, 0.1, 0.6}
\definecolor{commentcolor}{rgb}{0.4, 0.4, 0.4}
\definecolor{symbolcolor}{rgb}{0.0, 0.1, 0.6}
\definecolor{sortcolor}{rgb}{0.1, 0.5, 0.1}
\definecolor{attributecolor}{rgb}{0.7, 0.1, 0.1}
\lstdefinelanguage{lean}{
  mathescape=false,
  texcl=false,
  morekeywords=[1]{
    import, prelude, protected, private, noncomputable, definition, meta, renaming,
    hiding, parameter, parameters, begin, constant, constants,
    lemma, variable, variables, theory,
    print, theorem, example,
    open, as, export, override, axiom, axioms, inductive, with,
    structure, record, universe, universes,
    alias, help, precedence, reserve, declare_trace, add_key_equivalence,
    match, infix, infixl, infixr, notation, postfix, prefix, instance, reduce, check, end, this,
    using, using_well_founded, namespace, section,
    attribute, local, set_option, extends, include, omit, class,
    raw, replacing,
    calc, have, show, suffices, by, in, at, let, forall, Pi, fun,
    exists, if, dif, then, else, assume, obtain, from, register_simp_ext, unless, break, continue,
    mutual, do, def, run_cmd, const,
    partial, mut, where, macro, syntax, deriving,
    return, try, catch, for, macro_rules, declare_syntax_cat, abbrev
  },
  morekeywords=[2]{Sort, Type, Prop},
  morekeywords=[3]{
    assumption,
    apply, intro, intros, allGoals,
    generalize, clear, revert, done, exact,
    refine, repeat, cases, rewrite, rw,
    simp, simp_all, contradiction,
    constructor, injection,
    induction
  },
  literate=
    {α}{{\ensuremath{\alpha}}}1
    {β}{{\ensuremath{\beta}}}1
    {γ}{{\ensuremath{\gamma}}}1
    {δ}{{\ensuremath{\delta}}}1
    {ε}{{\ensuremath{\varepsilon}}}1
    {ζ}{{\ensuremath{\zeta}}}1
    {η}{{\ensuremath{\eta}}}1
    {θ}{{\ensuremath{\theta}}}1
    {ι}{{\ensuremath{\iota}}}1
    {κ}{{\ensuremath{\kappa}}}1
    {μ}{{\ensuremath{\mu}}}1
    {ν}{{\ensuremath{\nu}}}1
    {ξ}{{\ensuremath{\xi}}}1
    {π}{{\ensuremath{\pi}}}1
    {ρ}{{\ensuremath{\rho}}}1
    {σ}{{\ensuremath{\sigma}}}1
    {τ}{{\ensuremath{\tau}}}1
    {φ}{{\ensuremath{\varphi}}}1
    {χ}{{\ensuremath{\chi}}}1
    {ψ}{{\ensuremath{\psi}}}1
    {ω}{{\ensuremath{\omega}}}1
    {Γ}{{\ensuremath{\Gamma}}}1
    {Δ}{{\ensuremath{\Delta}}}1
    {Θ}{{\ensuremath{\Theta}}}1
    {Λ}{{\ensuremath{\Lambda}}}1
    {Σ}{{\color{symbolcolor}\ensuremath{\Sigma}}}1
    {Φ}{{\ensuremath{\Phi}}}1
    {Ξ}{{\ensuremath{\Xi}}}1
    {Ψ}{{\ensuremath{\Psi}}}1
    {Ω}{{\ensuremath{\Omega}}}1
    {Π}{{\color{symbolcolor}\ensuremath{\Pi}}}1
    {∀}{{\color{symbolcolor}\ensuremath{\forall}}}1
    {∃}{{\color{symbolcolor}\ensuremath{\exists}}}1
    {λ}{{\color{symbolcolor}\ensuremath{\lambda}}}1
    {ℵ}{{\ensuremath{\aleph}}}1
    {≤}{{\ensuremath{\leq}}}1
    {≥}{{\ensuremath{\geq}}}1
    {≠}{{\ensuremath{\neq}}}1
    {≈}{{\ensuremath{\approx}}}1
    {≡}{{\ensuremath{\equiv}}}1
    {≃}{{\ensuremath{\simeq}}}1
    {∂}{{\ensuremath{\partial}}}1
    {∆}{{\ensuremath{\triangle}}}1
    {∫}{{\ensuremath{\int}}}1
    {∑}{{\color{symbolcolor}\ensuremath{\sum}}}1
    {∏}{{\ensuremath{\prod}}}1
    {⊥}{{\ensuremath{\perp}}}1
    {∞}{{\ensuremath{\infty}}}1
    {∓}{{\ensuremath{\mp}}}1
    {±}{{\ensuremath{\pm}}}1
    {×}{{\ensuremath{\times}}}1
    {⊕}{{\ensuremath{\oplus}}}1
    {⊗}{{\ensuremath{\otimes}}}1
    {⊞}{{\ensuremath{\boxplus}}}1
    {∇}{{\ensuremath{\nabla}}}1
    {√}{{\ensuremath{\sqrt{}}}}1
    {⬝}{{\ensuremath{\cdot}}}1
    {•}{{\ensuremath{\cdot}}}1
    {∘}{{\ensuremath{\circ}}}1
    {·}{{\ensuremath{\cdot}}}1
    {⁻}{{\ensuremath{^{-}}}}1
    {▸}{{\ensuremath{\triangleright}}}1
    {∧}{{\ensuremath{\wedge}}}1
    {∨}{{\ensuremath{\vee}}}1
    {¬}{{\ensuremath{\neg}}}1
    {⊢}{{\ensuremath{\vdash}}}1
    {⟨}{{\ensuremath{\langle}}}1
    {⟩}{{\ensuremath{\rangle}}}1
    {↦}{{\ensuremath{\mapsto}}}1
    {←}{{\ensuremath{\leftarrow}}}1
    {→}{{\ensuremath{\rightarrow}}}1
    {↔}{{\ensuremath{\leftrightarrow}}}1
    {⇒}{{\ensuremath{\Rightarrow}}}1
    {⟹}{{\ensuremath{\Longrightarrow}}}1
    {⇐}{{\ensuremath{\Leftarrow}}}1
    {⟸}{{\ensuremath{\Longleftarrow}}}1
    {⊓}{{\ensuremath{\sqcap}}}1
    {⊔}{{\ensuremath{\sqcup}}}1
    {∩}{{\ensuremath{\cap}}}1
    {∪}{{\ensuremath{\cup}}}1
    {⊂}{{\ensuremath{\subseteq}}}1
    {⊆}{{\ensuremath{\subseteq}}}1
    {⊄}{{\ensuremath{\nsubseteq}}}1
    {⊈}{{\ensuremath{\nsubseteq}}}1
    {⊃}{{\ensuremath{\supseteq}}}1
    {⊇}{{\ensuremath{\supseteq}}}1
    {⊅}{{\ensuremath{\nsupseteq}}}1
    {⊉}{{\ensuremath{\nsupseteq}}}1
    {∈}{{\ensuremath{\in}}}1
    {∉}{{\ensuremath{\notin}}}1
    {∋}{{\ensuremath{\ni}}}1
    {∌}{{\ensuremath{\notni}}}1
    {∅}{{\ensuremath{\emptyset}}}1
    {∖}{{\ensuremath{\setminus}}}1
    {†}{{\ensuremath{\dag}}}1
    {ℕ}{{\ensuremath{\mathbb{N}}}}1
    {ℤ}{{\ensuremath{\mathbb{Z}}}}1
    {ℝ}{{\ensuremath{\mathbb{R}}}}1
    {ℚ}{{\ensuremath{\mathbb{Q}}}}1
    {ℂ}{{\ensuremath{\mathbb{C}}}}1
    {⌞}{{\ensuremath{\llcorner}}}1
    {⌟}{{\ensuremath{\lrcorner}}}1
    {⦃}{{\ensuremath{\{\!|}}}1
    {⦄}{{\ensuremath{|\!\}}}}1
    {‖}{{\ensuremath{\|}}}1
    {₀}{{\ensuremath{_0}}}1
    {₁}{{\ensuremath{_1}}}1
    {₂}{{\ensuremath{_2}}}1
    {₃}{{\ensuremath{_3}}}1
    {₄}{{\ensuremath{_4}}}1
    {₅}{{\ensuremath{_5}}}1
    {₆}{{\ensuremath{_6}}}1
    {₇}{{\ensuremath{_7}}}1
    {₈}{{\ensuremath{_8}}}1
    {₉}{{\ensuremath{_9}}}1
    {ᵢ}{{\ensuremath{_i}}}1
    {ⱼ}{{\ensuremath{_j}}}1
    {ₐ}{{\ensuremath{_a}}}1
    {ₙ}{{\ensuremath{_n}}}1
    {ₘ}{{\ensuremath{_m}}}1
    {ₚ}{{\ensuremath{_p}}}1
    {ₗ}{{\textsubscript{l}}}1
    {¹}{{\ensuremath{^1}}}1
    {↑}{{\ensuremath{\uparrow}}}1
    {↓}{{\ensuremath{\downarrow}}}1
    {...}{{\ensuremath{\ldots}}}1
    {:=}{{\color{symbolcolor}:=}}2
    {=}{{\color{symbolcolor}=}}1
    {<|>}{{\color{symbolcolor}<|>}}3
    {<\$>}{{\color{symbolcolor}<\$>}}3
    {+}{{\color{symbolcolor}+}}1
    {*}{{\color{symbolcolor}*}}1
    {\$}{{\color{symbolcolor}\$}}1,
  morecomment=[s][\color{commentcolor}]{/-}{-/},
  morecomment=[l][\itshape\color{commentcolor}]{--},
  morestring=[b]",
  showstringspaces=false,
  keepspaces=true,
  tabsize=2,
  sensitive=true,
  breaklines=true,
  breakatwhitespace=true,
  basicstyle=\ttfamily\small,
  captionpos=b,
  columns=[l]fullflexible,
  identifierstyle={\ttfamily\color{black}},
  keywordstyle=[1]{\ttfamily\color{keywordcolor}},
  keywordstyle=[2]{\ttfamily\color{sortcolor}},
  keywordstyle=[3]{\ttfamily\color{tacticcolor}},
  keywordstyle=[4]{\ttfamily\color{attributecolor}},
  stringstyle=\ttfamily,
  commentstyle={\ttfamily\footnotesize}
}
\lstdefinestyle{leanstyle}{
  language=lean,
  backgroundcolor=\color{lightblue},
  frame=single,
  framerule=0pt,
  xleftmargin=\fboxsep,
  xrightmargin=\fboxsep,
  inputencoding=utf8,
  extendedchars=true,
  columns=fullflexible,
  keepspaces=true
}

\DeclareUnicodeCharacter{2200}{$\forall$}
\DeclareUnicodeCharacter{220F}{$\prod$}
\DeclareUnicodeCharacter{2211}{$\sum$}
\DeclareUnicodeCharacter{2203}{$\exists$}
\DeclareUnicodeCharacter{2192}{$\rightarrow$}
\DeclareUnicodeCharacter{2265}{$\geq$}
\DeclareUnicodeCharacter{03A0}{\ensuremath{\Pi}}
\DeclareUnicodeCharacter{2097}{\textsubscript{l}}
\DeclareUnicodeCharacter{2019}{'}
\DeclareUnicodeCharacter{2013}{--}
\DeclareUnicodeCharacter{2014}{---}

\newtheorem{theorem}{Theorem}[section]
\newtheorem{lemma}[theorem]{Lemma}

\theoremstyle{definition}
\newtheorem{definition}[theorem]{Definition}

\definecolor{lightblue}{rgb}{0.8,0.85,1}
\definecolor{keywordcolor}{rgb}{0.7, 0.1, 0.1}
\definecolor{tacticcolor}{rgb}{0.0, 0.1, 0.6}
\definecolor{commentcolor}{rgb}{0.4, 0.4, 0.4}
\definecolor{symbolcolor}{rgb}{0.0, 0.1, 0.6}
\definecolor{sortcolor}{rgb}{0.1, 0.5, 0.1}
\definecolor{attributecolor}{rgb}{0.7, 0.1, 0.1}

\newcommand{\extlink}{~\ensuremath{{}^\text{\faExternalLink*}}}
\icmltitlerunning{Formalizing Scarf, Brouwer, and Nash in Lean}

\begin{document}

\twocolumn[
\icmltitle{Formalizing Scarf, Brouwer, and Nash in Lean}

\begin{icmlauthorlist}
\icmlauthor{Yuwei Lyu}{xmu}
\icmlauthor{Kai Li}{xmu}
\end{icmlauthorlist}

\icmlaffiliation{xmu}{Department of Mathematics, Xiamen University Malaysia, Sepang, Selangor, Malaysia}

\icmlcorrespondingauthor{Yuwei Lyu}{Lyuyuwei@gmail.com}
\icmlcorrespondingauthor{Kai Li}{donk28510@gmail.com}

\vskip 0.3in
]

\printAffiliationsAndNotice{}

\begin{abstract}
We formalize in Lean 4 a complete combinatorial route from Scarf's theorem to Brouwer's fixed point theorem and to the existence of mixed Nash equilibria in finite games.  The development follows Ivanov's indexed-order formulation of Scarf's theorem, formalizes the room--door incidence structure and parity argument, instantiates the theorem on finite grids of the standard simplex, and carries out the compactness and continuity argument needed to obtain a fixed point.  We then extend the result to finite products of simplices by an explicit embedding--projection construction and use this product theorem to prove mixed Nash equilibrium existence via the Nash map.  As a secondary by-product, we derive BrouwerBench, a preliminary 80-item Lean-grounded benchmark for probing proof-structure understanding within this single formal development.
\end{abstract}

\section{Introduction}

Brouwer's fixed point theorem is a fundamental existence result in topology~\cite{brouwer1911beweis} and a standard tool in analysis, economics, and game theory~\cite{border1985fixed}.  A central application is Nash's theorem for finite games~\cite{nash1950equilibrium}: mixed strategy profiles form a finite product of simplices, and an appropriately defined continuous self-map has a fixed point exactly at a mixed Nash equilibrium.  For formalization, it is useful not only to invoke Brouwer's theorem as a black-box topological principle, but also to expose the finite combinatorial structure from which one can derive the fixed-point result.  This makes the proof pipeline modular: the finite parity argument, the limiting construction, the product-of-simplices reduction, and the game-theoretic endpoint can be checked as separate formal components.

This paper formalizes such a route in Lean 4.  We start from Ivanov's indexed-order formulation of Scarf's combinatorial theorem~\cite{scarf1982computation,ivanov2019beyond}, where a finite set is equipped with indexed linear orders and a coloring.  The proof builds a finite incidence structure of dominant sets, cells, rooms, and doors: outside doors are incident to one room, internal doors are incident to two rooms, and the resulting parity argument produces a colorful room.  We then instantiate this theorem on increasingly fine grids of the standard simplex.  The colorful rooms provide discrete approximations, while compactness, continuity, and vanishing-diameter estimates yield a Brouwer fixed point.  Finally, we extend the standard-simplex theorem to finite products of simplices by an explicit embedding--projection construction and apply the product theorem to the Nash map for finite games.

The formalization treats this route as a sequence of reusable proof components rather than only as a final fixed-point theorem.  In particular, the room--door parity argument, the dominance estimates on simplex grids, and the embedding--projection reduction for products of simplices are exposed as named Lean definitions and lemmas.  This makes the development inspectable as a structured proof artifact and allows the final Nash theorem to be traced back to finite combinatorial data.

The development builds on Mathlib~\cite{mathlib2020} and is written in Lean 4~\cite{moura2021lean}. Mathlib provides finite types, finite sets, real analysis, compactness, continuity, finite-dimensional spaces, and standard simplices.  The development consists of 4,768 lines of Lean across five files, covering the Scarf combinatorics, the standard-simplex Brouwer proof, the product-simplex reduction, the Nash endpoint, and supporting simplex infrastructure; Appendix~\ref{app:formalization-overview} gives the file-level breakdown.  The paper-specific Lean artifact is publicly available.\footnote{Code and benchmark materials are available in the \href{https://github.com/Solo-ary/Game-Theory-Formalization/tree/camera-ready-icml2026}{camera-ready artifact}; continued development is maintained in the \href{https://github.com/math-xmum/Brouwer}{upstream repository}.}

The body of the paper presents the proof pipeline and main formalization choices.  Appendices~\ref{app:scarf-details}--\ref{app:nash-details} summarize the main proof components, and Appendix~\ref{app:selected-interfaces} records representative Lean interfaces.  The clickable symbol~\faExternalLink*{} links mathematical statements to their full definitions and proofs in the source files.  Since each stage is represented by named Lean definitions and lemmas, the development also yields local proof-understanding tasks.  We package these tasks as BrouwerBench, a small pilot benchmark intended to probe whether models can answer proof-role questions inside this particular Lean-grounded development.  BrouwerBench should therefore be read as a secondary artifact of the formalization, not as a broad benchmark for general mathematical reasoning or Lean theorem proving.

\subsection{Contributions}
\label{sec:contributions}

Our main contribution is a Lean-checked proof pipeline from Scarf's
combinatorial theorem to Brouwer's fixed point theorem and mixed Nash
equilibrium existence.  Specifically, we formalize Ivanov's indexed-order
version of Scarf's theorem, including dominance, rooms, doors, the two-room
property, and the parity argument for colorful rooms; instantiate this theorem
on finite simplex grids to derive Brouwer's theorem on the standard simplex by
dominance estimates, compactness, and continuity; extend the result to finite
products of simplices using an explicit embedding--projection construction;
and apply the product theorem to prove mixed Nash equilibrium existence for
finite games.  As a secondary by-product, we extract BrouwerBench, an 80-item
Lean-grounded pilot benchmark for proof-structure understanding within this
formalization.

\section{Related Work}
\label{sec:related}

\paragraph{Formalized mathematics in Lean.}
Lean and Mathlib have supported increasingly large formalized mathematics
projects, from undergraduate algebra and topology to research-level
developments~\cite{mathlib2020,moura2021lean}.  These projects show that
modern proof assistants can serve not only as proof checkers, but also as
environments for organizing reusable mathematical infrastructure.  Our work
follows this line, but focuses on a proof route whose structure combines finite
combinatorics with analysis: a Scarf-style parity argument is formalized first,
then transported to Brouwer's fixed point theorem by approximation and
compactness, and finally applied to Nash equilibria.

\paragraph{Fixed-point and equilibrium theorems.}
Brouwer's fixed point theorem is a classical result with many proofs and
applications, including the standard route to the existence of mixed Nash
equilibria in finite games~\cite{brouwer1911beweis,nash1950equilibrium,border1985fixed}.
The emphasis of our development is not only the final fixed-point or
equilibrium statement, but the formalized proof pipeline leading to it.  We
formalize Scarf's combinatorial theorem~\cite{scarf1982computation,ivanov2019beyond},
derive Brouwer's theorem on the standard simplex, extend it to finite products
of simplices, and then obtain Nash equilibria through the Nash map.  This
makes the finite room--door incidence argument, dominance estimates on simplex
grids, limiting construction, and product-simplex reduction explicit as
reusable Lean components, rather than treating Brouwer's theorem as a black
box.

\paragraph{Autoformalization and benchmarks.}
Recent work on neural theorem proving and autoformalization studies how
language models translate mathematical text into formal statements and proofs,
retrieve useful lemmas, and construct or repair proof scripts~\cite{polu2020generative,zheng2021minif2f}.
Many benchmarks in this area consist of independent problem instances or
theorem statements.  BrouwerBench is instead extracted from a single
kernel-checked Lean development: each item is tied to a concrete definition,
dependency, lemma role, or design decision in the Scarf--Brouwer--Nash
pipeline.  It is intended as a small pilot probe of proof-structure
understanding inside this coherent formalization, rather than as a broad
model-ranking benchmark for general theorem proving or mathematical reasoning.

\section{Formalizing Scarf's Combinatorial Theorem}\label{sec:Scarf}
This section contains the finite combinatorial part of the development. We follow Ivanov's abstract formulation of Scarf's theorem~\cite{ivanov2019beyond}, where the input is a finite set equipped with a family of indexed linear orders and a coloring. The proof is organized around three layers. First, dominance turns order-theoretic data into finite cells. Second, rooms and doors provide the incidence structure used for counting. Third, colorings and nearly colorful cells turn this incidence structure into a parity argument producing a colorful room.

\subsection{Orders, Dominant Sets, and Cells}\label{subsec:scarf-orders}

We work with a finite type $T$ together with a finite index type $I$. For each $i \in I$, we are given a linear order $<_i$ on $T$. The indexed family of orders is the basic object from which dominance is defined. In Lean, this data is bundled as a typeclass\href{https://github.com/Solo-ary/Game-Theory-Formalization/blob/camera-ready-icml2026/Brouwer/Gametheory/Scarf.lean#L88-L93}{\extlink}, so that the notation for comparisons can depend on the index.

\begin{definition}[Indexed family of linear orders]\label{def:linearorder}
A family of linear orders on $T$ indexed by $I$ is a collection $(<_i)_{i \in I}$ such that each $<_i$ is a linear order on $T$\href{https://github.com/Solo-ary/Game-Theory-Formalization/blob/camera-ready-icml2026/Brouwer/Gametheory/Scarf.lean#L88-L89}{\extlink}.
\end{definition}

Given a nonempty subset $\sigma \subseteq T$ and an index $i \in I$, we write $\min_i \sigma$ for the minimum of $\sigma$ with respect to $<_i$. Dominance says that every point of $T$ is controlled by at least one of the orders indexed by $C$. This is the order-theoretic condition that later forces the relevant finite sets to have the correct cardinalities.

\begin{definition}[Dominant set]
Let $C$ be a nonempty subset of $I$. A subset $\sigma$ of $T$ is dominant with respect to $C$ if, for every element $y$ of $T$, one can find an index $i$ in $C$ such that $y$ is below every element of $\sigma$ in the order $<_i$\href{https://github.com/Solo-ary/Game-Theory-Formalization/blob/camera-ready-icml2026/Brouwer/Gametheory/Scarf.lean#L106-L108}{\extlink}.

When $\sigma$ is nonempty, this is equivalent to saying that no element of $T$ is strictly larger than all the minima $\min_i \sigma$ for $i$ in $C$.
\end{definition}

We adopt the convention that $\emptyset \subset T$ is dominant with respect to every non-empty subset $C \subset I$\href{https://github.com/Solo-ary/Game-Theory-Formalization/blob/camera-ready-icml2026/Brouwer/Gametheory/Scarf.lean#L167-L175}{\extlink}. The following lemma collects the basic closure and cardinality facts about dominant sets. These facts are used repeatedly when passing between cells, rooms, and doors.

\begin{lemma}[Basic properties of dominant sets]\label{def:Lemma1}
If $\sigma$ is nonempty and dominant with respect to $C$, then
$\sigma = \{\min_i \sigma \mid i \in C\}$, and hence
$|\sigma| \le |C|$\href{https://github.com/Solo-ary/Game-Theory-Formalization/blob/camera-ready-icml2026/Brouwer/Gametheory/Scarf.lean#L138-L165}{\extlink}.
If $\tau \subset \sigma$, then $\tau$ is dominant with respect to
$C$\href{https://github.com/Solo-ary/Game-Theory-Formalization/blob/camera-ready-icml2026/Brouwer/Gametheory/Scarf.lean#L115-L124}{\extlink},
and if $C \subset D$, then $\sigma$ is dominant with respect to
$D$\href{https://github.com/Solo-ary/Game-Theory-Formalization/blob/camera-ready-icml2026/Brouwer/Gametheory/Scarf.lean#L126-L134}{\extlink}.
\end{lemma}

The preceding lemma allows us to treat dominant pairs $(\sigma,C)$ as cells of a finite combinatorial object. Rooms are the full-dimensional cells, while doors are their codimension-one boundaries in the counting argument.

\begin{definition}[Cells, rooms, and doors]\label{def:Cells, rooms, and doors}
A \emph{cell} is a pair $(\sigma, C)$ where $\sigma$ is dominant with respect to $C$. A cell is called a \emph{room} if $|C| = |\sigma|$\href{https://github.com/Solo-ary/Game-Theory-Formalization/blob/camera-ready-icml2026/Brouwer/Gametheory/Scarf.lean#L177-L179}{\extlink}, and a \emph{door} if $|C| = |\sigma| + 1$\href{https://github.com/Solo-ary/Game-Theory-Formalization/blob/camera-ready-icml2026/Brouwer/Gametheory/Scarf.lean#L188}{\extlink}.
\end{definition}

\subsection{Rooms, Doors, and the Two-Room Property}\label{subsec:scarf-doors}

The next step is to make precise how rooms meet along doors. In the formalization, this is captured by the predicate \verb|isDoorof|\href{https://github.com/Solo-ary/Game-Theory-Formalization/blob/camera-ready-icml2026/Brouwer/Gametheory/Scarf.lean#L193-L195}{\extlink}, which records that a door is obtained from a room either by removing an element from the underlying set or by adding a color. This relation is the local incidence relation used later in the parity count.

\begin{definition}[Outside and internal doors]\label{def:Outside and internal doors}
A door $(\tau,D)$ is called an \emph{outside door} if $\tau = \emptyset$, and an \emph{internal door} otherwise\href{https://github.com/Solo-ary/Game-Theory-Formalization/blob/camera-ready-icml2026/Brouwer/Gametheory/Scarf.lean#L252-L256}{\extlink}.
\end{definition}

The following two lemmas give the degree structure of this incidence relation. Outside doors behave as boundary objects: they have degree one. Internal doors behave as shared faces: they have degree two.

\begin{lemma}[Outside-door property]\label{lemma:outsidedoor-property}
Suppose that $|D| = 1$. Then $(\varnothing, D)$ is an outside door\href{https://github.com/Solo-ary/Game-Theory-Formalization/blob/camera-ready-icml2026/Brouwer/Gametheory/Scarf.lean#L258-L271}{\extlink} and is a door of exactly one room. Every outside door has this form\href{https://github.com/Solo-ary/Game-Theory-Formalization/blob/camera-ready-icml2026/Brouwer/Gametheory/Scarf.lean#L274-L283}{\extlink}.
\end{lemma}

\begin{lemma}[Two-room property]\label{lemma:two-room-property}
Every internal door belongs to exactly two rooms\href{https://github.com/Solo-ary/Game-Theory-Formalization/blob/camera-ready-icml2026/Brouwer/Gametheory/Scarf.lean#L760-L1148}{\extlink}.
\end{lemma}

Thus the rooms and doors form a finite graph-like incidence structure: starting from an outside door, one enters a unique room, and every time one crosses an internal door there is a unique other room on the opposite side. This is the combinatorial shape behind the later path-following and parity argument.

The proof of the two-room property is the most involved local part of the Scarf formalization. It analyzes an internal door $(\tau, D)$ via two complementary constructions: either removing a color from $D$\href{https://github.com/Solo-ary/Game-Theory-Formalization/blob/camera-ready-icml2026/Brouwer/Gametheory/Scarf.lean#L318-L394}{\extlink}, or extending $\tau$ by a suitable element\href{https://github.com/Solo-ary/Game-Theory-Formalization/blob/camera-ready-icml2026/Brouwer/Gametheory/Scarf.lean#L396-L595}{\extlink}. These are implemented in Lean via the auxiliary constructions \verb|M_set|\href{https://github.com/Solo-ary/Game-Theory-Formalization/blob/camera-ready-icml2026/Brouwer/Gametheory/Scarf.lean#L289-L291}{\extlink} and maximal elements \verb|m_element|\href{https://github.com/Solo-ary/Game-Theory-Formalization/blob/camera-ready-icml2026/Brouwer/Gametheory/Scarf.lean#L297-L316}{\extlink}, which ensure that both constructions yield valid rooms. In the paper, we use the two-room property as a black-box incidence theorem; in Lean, the proof supplies the exact finite-set witnesses needed to make the incidence count well typed.

\subsection{Colorings and the Parity Argument}\label{subsec:scarf-colorings}

We now add the coloring data. The coloring selects, for each element of $T$, an index in $I$. A colorful cell is one whose index set is exactly the image of its underlying set under this coloring. Nearly colorful cells differ from being colorful by exactly one missing color; the missing color is recorded as the type. This typed version is useful because the parity argument counts objects of a fixed type.

\begin{definition}[Coloring]\label{def:coloring}
Fix a coloring $c:T\to I$. A cell $(\sigma,C)$ is called \emph{colorful} if
$C=c(\sigma),$
and \emph{nearly colorful} if
$|C\setminus c(\sigma)|=1.$
For a nearly colorful cell $(\sigma,C)$, the unique element of $C\setminus c(\sigma)$ is called its \emph{type}.
\end{definition}

The next lemmas connect the uncolored incidence structure with the coloring data. They identify the boundary terms, describe how colorful rooms contribute nearly colorful doors, and show that nearly colorful rooms have the expected number of nearly colorful doors.

\begin{lemma}[Outside doors and doors of colorful rooms]\label{lemma:4}
For every $i \in I$ there is exactly one outside door of the type $i$. Every door of a
colorful room is nearly colorful and there is exactly one door of each type among them.
\end{lemma}

\begin{lemma}[Nearly colorful door transition]\label{lemma:5}
Suppose that $(\tau, D)$ is a nearly colorful door of a room $(\sigma, C)$. Then $(\sigma, C)$
is either colorful, or nearly colorful of the same type as $(\tau, D)$\href{https://github.com/Solo-ary/Game-Theory-Formalization/blob/camera-ready-icml2026/Brouwer/Gametheory/Scarf.lean#L1263-L1341}{\extlink}.
\end{lemma}

\begin{lemma}[Nearly colorful cardinality properties]\label{lemma:6}
If $(\sigma, C)$ is a nearly colorful room, then $|c(\sigma)|$ is equal either to $|\sigma|$ or to
$|\sigma| - 1$\href{https://github.com/Solo-ary/Game-Theory-Formalization/blob/camera-ready-icml2026/Brouwer/Gametheory/Scarf.lean#L1370-L1394}{\extlink}. If $(\sigma, C)$ is a nearly colorful door, then $c(\sigma) \subset C$\href{https://github.com/Solo-ary/Game-Theory-Formalization/blob/camera-ready-icml2026/Brouwer/Gametheory/Scarf.lean#L1396-L1420}{\extlink}.
\end{lemma}

\begin{lemma}[Doors of nearly colorful rooms]\label{lemma:7}
If $(\sigma, C)$ is a nearly colorful room, then $(\sigma, C)$ has two nearly colorful doors\href{https://github.com/Solo-ary/Game-Theory-Formalization/blob/camera-ready-icml2026/Brouwer/Gametheory/Scarf.lean#L1562-L2211}{\extlink}.
\end{lemma}

Together, these lemmas turn the room--door incidence relation into a parity count. Fixing a type $i$, the unique outside door of type $i$ contributes one boundary term\href{https://github.com/Solo-ary/Game-Theory-Formalization/blob/camera-ready-icml2026/Brouwer/Gametheory/Scarf.lean#L2223-L2343}{\extlink}. Internal nearly colorful doors are counted twice because of the two-room property\href{https://github.com/Solo-ary/Game-Theory-Formalization/blob/camera-ready-icml2026/Brouwer/Gametheory/Scarf.lean#L2345-L2457}{\extlink}. Nearly colorful rooms also contribute an even number of typed doors\href{https://github.com/Solo-ary/Game-Theory-Formalization/blob/camera-ready-icml2026/Brouwer/Gametheory/Scarf.lean#L2470-L2562}{\extlink}. Therefore the remaining contribution comes from colorful rooms, forcing the relevant colorful-room count to be odd\href{https://github.com/Solo-ary/Game-Theory-Formalization/blob/camera-ready-icml2026/Brouwer/Gametheory/Scarf.lean#L2579-L2585}{\extlink}.

\begin{theorem}[Scarf's combinatorial theorem]\label{thm:Scarf}
For every coloring $c : T \to I$, there exists a colorful room. Moreover, the number of colorful rooms is odd\href{https://github.com/Solo-ary/Game-Theory-Formalization/blob/camera-ready-icml2026/Brouwer/Gametheory/Scarf.lean#L2587-L2595}{\extlink}.
\end{theorem}

This theorem completes the finite combinatorial core of the formalization. In the next section, the finite set $T$ will be instantiated by a grid in the standard simplex. The colorful rooms supplied by Scarf's theorem then become the discrete approximations from which Brouwer fixed points are obtained by a limiting argument.

\section{From Scarf's Theorem to Brouwer's Fixed Point Theorem}\label{sec:Brouwer}

This section derives Brouwer's fixed point theorem from the formalized Scarf theorem. The argument has two stages. We first prove Brouwer's theorem for the standard simplex by applying Scarf's theorem to increasingly fine finite grids and then passing to a limit. We then obtain the corresponding theorem for finite products of simplices by reducing the product case to the standard-simplex case through an explicit embedding--projection construction.

\subsection{Brouwer's Theorem on the Standard Simplex}\label{subsec:brouwer-standard}

Fix a positive integer $n$, and let $\Delta^{n-1}$ be the standard simplex with coordinates indexed by $\mathrm{Fin}\ n$. The goal is to prove that every continuous self-map
$$
f : \Delta^{n-1} \to \Delta^{n-1}
$$
has a fixed point. The proof does not use Scarf's theorem directly on the simplex itself. Instead, for each positive integer $l$, we work with a finite grid $T_l$ of rational points whose coordinates have denominator $l$. Informally,
$$
T_l =
\left\{
  x \in \Delta^{n-1}
  \mid
  x_i \in \frac{1}{l}\mathbb{Z}
  \text{ for all } i
\right\}.
$$
In Lean, the grid is represented by integer coordinates in \verb|Fin (l+1)| whose sum is $l$; the corresponding point of the simplex is obtained by dividing each coordinate by $l$. The subtype \verb|TT| defines the finite grid, while \verb|TTtostdSimplex| and the coercion instance \verb|TT.CoestdSimplex| realize a grid point as an element of the standard simplex\href{https://github.com/Solo-ary/Game-Theory-Formalization/blob/camera-ready-icml2026/Brouwer/Gametheory/Brouwer.lean#L16-L50}{\extlink}.

\subsubsection{The ordered grid and dominance estimates}

To apply Scarf's theorem, each grid $T_l$ is equipped with an indexed family of linear orders, one order for each coordinate. For a fixed coordinate $i$, the order first compares the $i$-th coordinate and then uses a lexicographic tie-breaker on the whole grid point. This is encoded by \verb|TT.Ilt| and assembled into the indexed linear order \verb|TT.ILO|\href{https://github.com/Solo-ary/Game-Theory-Formalization/blob/camera-ready-icml2026/Brouwer/Gametheory/Brouwer.lean#L53-L76}{\extlink}. These orders are chosen so that dominance in the sense of Scarf's theorem imposes strong metric restrictions on the corresponding grid points. This is the point at which the abstract combinatorial theorem becomes a quantitative statement about the simplex.

The key quantitative input is that dominance on the grid gives metric control. A dominant set has small diameter, and the coordinates outside its color set are uniformly small. These estimates are the bridge between the finite combinatorial theorem and the limiting argument.

\begin{theorem}[Dominance estimates on the grid]
Let $\sigma \subseteq T_l$ and $C \subseteq \mathrm{Fin}\ n$. If $\sigma$ is dominant with respect to $C$, then
$$
|x_i-y_i| < \frac{2(n+1)}{l}
$$
for every $x,y\in \sigma$ and every $i \in \mathrm{Fin}\ n$
\href{https://github.com/Solo-ary/Game-Theory-Formalization/blob/camera-ready-icml2026/Brouwer/Gametheory/Brouwer.lean#L203-L315}{\extlink}.
Moreover, if $i\notin C$, then
$$
x_i < \frac{n+1}{l}
$$
for every $x\in\sigma$
\href{https://github.com/Solo-ary/Game-Theory-Formalization/blob/camera-ready-icml2026/Brouwer/Gametheory/Brouwer.lean#L317-L362}{\extlink}.
\end{theorem}

Thus, as $l$ tends to infinity, dominant rooms become small, and the coordinates not belonging to their color sets vanish in the limit.

\subsubsection{The coloring induced by a self-map}

Let $f:\Delta^{n-1}\to\Delta^{n-1}$ be continuous. For any point $x\in\Delta^{n-1}$, the coordinate sums of $x$ and $f(x)$ are both equal to $1$. Hence not all coordinates can strictly decrease: there exists an index $i$ such that
$$
x_i \leq f(x)_i.
$$
The formalization packages this elementary observation as the nonemptiness lemma \verb|stdSimplex.upidx|\href{https://github.com/Solo-ary/Game-Theory-Formalization/blob/camera-ready-icml2026/Brouwer/Gametheory/Brouwer.lean#L372-L389}{\extlink} and the choice function \verb|stdSimplex.pick|\href{https://github.com/Solo-ary/Game-Theory-Formalization/blob/camera-ready-icml2026/Brouwer/Gametheory/Brouwer.lean#L391}{\extlink} on the simplex. Restricting this choice to the grid $T_l$ gives the coloring required by Scarf's theorem, namely \verb|Fcolor|\href{https://github.com/Solo-ary/Game-Theory-Formalization/blob/camera-ready-icml2026/Brouwer/Gametheory/Brouwer.lean#L395-L395}{\extlink}.

Applying the formalized Scarf theorem to \verb|Fcolor| produces a colorful room on each grid $T_l$. In Lean, these rooms are chosen as a sequence \verb|room_seq|\href{https://github.com/Solo-ary/Game-Theory-Formalization/blob/camera-ready-icml2026/Brouwer/Gametheory/Brouwer.lean#L397-L399}{\extlink}, and a representative point from each room is chosen as \verb|room_point_seq|\href{https://github.com/Solo-ary/Game-Theory-Formalization/blob/camera-ready-icml2026/Brouwer/Gametheory/Brouwer.lean#L401-L402}{\extlink}. This sequence of rooms is the bridge between the finite combinatorial theorem and the limiting proof of Brouwer's theorem.

\subsubsection{Compactness and the limiting argument}

Choose one representative point from the colorful room obtained at each grid level. Since there are only finitely many possible color sets, we first pass to a subsequence on which the color set of the chosen colorful rooms is constant. This finite-image subsequence extraction is formalized by
\texttt{exists\_subseq\_constant\_of\_finite\_image}
\href{https://github.com/Solo-ary/Game-Theory-Formalization/blob/camera-ready-icml2026/Brouwer/Gametheory/Brouwer.lean#L408-L414}{\extlink};
the resulting constant color set and order embedding are packaged as \verb|gpkg| and \verb|g1|
\href{https://github.com/Solo-ary/Game-Theory-Formalization/blob/camera-ready-icml2026/Brouwer/Gametheory/Brouwer.lean#L480-L489}{\extlink}.
We then use compactness of the standard simplex to extract a convergent subsequence of the corresponding room points; this step is packaged in Lean as \verb|hpkg_aux| and \verb|hpkg|
\href{https://github.com/Solo-ary/Game-Theory-Formalization/blob/camera-ready-icml2026/Brouwer/Gametheory/Brouwer.lean#L537-L548}{\extlink}.
Let $z\in\Delta^{n-1}$ denote the resulting limit, and write the constant color set as $C$.

The dominance estimates now control the limit. First, the diameters of the selected colorful rooms tend to zero\href{https://github.com/Solo-ary/Game-Theory-Formalization/blob/camera-ready-icml2026/Brouwer/Gametheory/Brouwer.lean#L556-L614}{\extlink}. Therefore every point in the same selected room converges to the same limit $z$. Second, for each coordinate $i\notin C$, the outside-coordinate estimate gives convergence to zero, and hence
$$
z_i=0
\qquad\text{for all } i\notin C\href{https://github.com/Solo-ary/Game-Theory-Formalization/blob/camera-ready-icml2026/Brouwer/Gametheory/Brouwer.lean#L497-L535}{\extlink}.
$$
Since $z$ lies in the simplex, it follows that
$$
\sum_{i\in C} z_i = 1.
$$
On the other hand, the coloring rule says that for each color $i\in C$, the chosen room contains grid points whose $i$-th coordinate does not decrease under $f$. Passing to the limit and using continuity of $f$ gives
$$
f(z)_i \ge z_i
\qquad\text{for all } i\in C\href{https://github.com/Solo-ary/Game-Theory-Formalization/blob/camera-ready-icml2026/Brouwer/Gametheory/Brouwer.lean#L616-L708}{\extlink}.
$$
Finally, $f(z)$ is also a point of the simplex. The inequalities on $C$, together with $\sum_{i\in C}z_i=1$, force equality on all coordinates in $C$ and force all coordinates outside $C$ to be zero. Hence $f(z)=z$.

\begin{theorem}[Brouwer's fixed point theorem for the standard simplex]
Every continuous map $f:\Delta^{n-1}\to\Delta^{n-1}$ has a fixed point\href{https://github.com/Solo-ary/Game-Theory-Formalization/blob/camera-ready-icml2026/Brouwer/Gametheory/Brouwer.lean#L710-L791}{\extlink}.
\end{theorem}

\subsection{Brouwer's Theorem on Products of Simplices}
\label{subsec:brouwer-product}

We next extend Brouwer's fixed point theorem from the standard simplex to finite products of simplices. This is the form needed for the game-theoretic application, where a mixed strategy profile is a point in a product of simplices. The formal proof reduces the product case to the standard-simplex case by constructing an explicit embedding--projection pair between the product and a larger simplex.

\subsubsection{Reducing products to a standard simplex}

Let $I$ be a finite index type, and let $\mathrm{card}(i)$ be a positive integer for each $i\in I$. The product space considered in Lean is
$$
\prod_{i\in I}\Delta^{\mathrm{card}(i)-1}.
$$
It is represented by the type \texttt{ProductSimplices card}. The ambient simplex has total coordinate number
$$
N=\sum_{i\in I}\mathrm{card}(i),
$$
and is represented by \texttt{BigSimplex card}
\href{https://github.com/Solo-ary/Game-Theory-Formalization/blob/camera-ready-icml2026/Brouwer/Gametheory/Brouwer_product.lean#L8-L18}{\extlink}.

The main bookkeeping issue is that coordinates in the ambient simplex are indexed by a single flat index, while coordinates in the product are indexed by a pair consisting of a component $i\in I$ and a coordinate inside the $i$-th simplex. In Lean, this conversion is handled by \texttt{index\_split} and \texttt{index\_combine}
\href{https://github.com/Solo-ary/Game-Theory-Formalization/blob/camera-ready-icml2026/Brouwer/Gametheory/Brouwer_product.lean#L105-L203}{\extlink}.
These maps allow the formalization to move between the block decomposition of the product and the flat coordinate system of the larger simplex.

\subsubsection{The embedding-projection construction}

The embedding sends a point of the product of simplices to a point of the ambient simplex by placing each component simplex into its corresponding coordinate block. The $i$-th block is rescaled by the weight $\mathrm{card}(i)/N$, so that the total sum of all coordinates is one. This construction is formalized as \texttt{embed\_from\_product}
\href{https://github.com/Solo-ary/Game-Theory-Formalization/blob/camera-ready-icml2026/Brouwer/Gametheory/Brouwer_product.lean#L374-L411}{\extlink}.

The projection in the opposite direction is more delicate. Starting from a point of the ambient simplex, the natural idea is to recover each component simplex by normalizing the corresponding coordinate block. However, this normalization is not well-defined if one of the blocks has total mass zero. To avoid this degenerate case, the formal construction first pushes the ambient point toward a fixed uniform point \texttt{z\_uniform}. This auxiliary operation is implemented by \texttt{pushTowardsZ}, together with the definitions \texttt{deficit} and \texttt{tPush}
\href{https://github.com/Solo-ary/Game-Theory-Formalization/blob/camera-ready-icml2026/Brouwer/Gametheory/Brouwer_product.lean#L223-L292}{\extlink}.
The key property is that after this push, every coordinate block has positive total mass
\href{https://github.com/Solo-ary/Game-Theory-Formalization/blob/camera-ready-icml2026/Brouwer/Gametheory/Brouwer_product.lean#L525-L558}{\extlink}.
The projection is then obtained by normalizing each block, and is formalized as \texttt{project\_to\_product}
\href{https://github.com/Solo-ary/Game-Theory-Formalization/blob/camera-ready-icml2026/Brouwer/Gametheory/Brouwer_product.lean#L294-L372}{\extlink}.

The central property of the construction is the retraction identity
$$
\texttt{project\_to\_product}\circ \texttt{embed\_from\_product}
=
\mathrm{id}.
$$
For points already coming from the product, each coordinate block has the prescribed positive total mass, so the projection recovers the original product point. This identity is formalized by \texttt{project\_embed\_id}
\href{https://github.com/Solo-ary/Game-Theory-Formalization/blob/camera-ready-icml2026/Brouwer/Gametheory/Brouwer_product.lean#L413-L486}{\extlink}.

\subsubsection{The product fixed-point theorem}

It remains to transfer Brouwer's theorem from the ambient simplex back to the product. The embedding and projection are proved continuous as \texttt{embed\_continuous} and \texttt{project\_continuous}
\href{https://github.com/Solo-ary/Game-Theory-Formalization/blob/camera-ready-icml2026/Brouwer/Gametheory/Brouwer_product.lean#L560-L652}{\extlink}
.
Given a continuous map
$$
f:\prod_{i\in I}\Delta^{\mathrm{card}(i)-1}
\to
\prod_{i\in I}\Delta^{\mathrm{card}(i)-1},
$$
we form a continuous self-map of the ambient simplex by composing projection, $f$, and embedding:
$$
F =
\texttt{embed\_from\_product}\circ f\circ \texttt{project\_to\_product}.
$$
The standard-simplex version of Brouwer's theorem gives a fixed point of $F$. Projecting this fixed point back to the product and using the retraction identity gives a fixed point of $f$.

\begin{theorem}[Brouwer fixed point theorem for products of simplices]
Every continuous self-map of a finite product of simplices has a fixed point
\href{https://github.com/Solo-ary/Game-Theory-Formalization/blob/camera-ready-icml2026/Brouwer/Gametheory/Brouwer_product.lean#L654-L669}{\extlink}.
\end{theorem}

\section{Nash Equilibrium Existence in Finite Games}\label{sec:Nash}
The final layer of the development turns the product fixed-point theorem into the classical existence theorem for finite games.  This section is deliberately small: once Brouwer has been established for products of simplices, the remaining work is to define the mixed payoff function, construct the Nash self-map, and prove that fixed points of this map are exactly mixed equilibria.

\subsection{Finite Games and Mixed Profiles}
A strategic-form game is represented by a type of players $I$, a dependent family of pure strategy types $S^i$, and payoff functions $g^i : (\Pi_i S^i) \to \mathbb{R}$.  The base structure \texttt{Game} only assumes that the player and strategy types are inhabited; the finite assumptions are isolated in \texttt{FinGame}.\href{https://github.com/Solo-ary/Game-Theory-Formalization/blob/camera-ready-icml2026/Brouwer/Gametheory/Nash.lean#L16-L49}{\extlink}  For a finite game, the mixed strategy space is the dependent product
$$
  \Delta(G) = \prod_{i\in I} \Delta(S^i),
$$
implemented as \texttt{mixedS := (i : G.I) -> stdSimplex R (G.SS i)}.\href{https://github.com/Solo-ary/Game-Theory-Formalization/blob/camera-ready-icml2026/Brouwer/Gametheory/Nash.lean#L59-L60}{\extlink}

The expected payoff of player $i$ at a mixed profile $\sigma$ is the finite multilinear extension of the pure payoff function:
$$
  \bar g^i(\sigma)
  = \sum_{s \in \Pi_j S^j}
      \left(\prod_j \sigma^j(s^j)\right) g^i(s).
$$
In Lean this is \texttt{mixed\_g}.\href{https://github.com/Solo-ary/Game-Theory-Formalization/blob/camera-ready-icml2026/Brouwer/Gametheory/Nash.lean#L62}{\extlink}  The key algebraic lemma is that \texttt{mixed\_g} is affine in one player's mixed strategy when the other components are fixed:
$$
  \bar g^i(\sigma[i\leftarrow y])
  = \sum_{a\in S^i} y(a)\,\bar g^i(\sigma[i\leftarrow \delta_a]).
$$
This lemma is the bridge from pure deviations to arbitrary mixed deviations in the final equilibrium proof.\href{https://github.com/Solo-ary/Game-Theory-Formalization/blob/camera-ready-icml2026/Brouwer/Gametheory/Nash.lean#L67-L134}{\extlink}

A mixed Nash equilibrium is a mixed profile $\sigma^*$ such that no player can improve expected payoff by replacing only their own component:
$$
  \forall i,\ \forall y\in\Delta(S^i),\qquad
  \bar g^i(\sigma^*) \geq \bar g^i(\sigma^*[i\leftarrow y]).
$$
This is the predicate \texttt{mixedNashEquilibrium}.\href{https://github.com/Solo-ary/Game-Theory-Formalization/blob/camera-ready-icml2026/Brouwer/Gametheory/Nash.lean#L165-L167}{\extlink}

\subsection{The Nash Map}
For a mixed profile $\sigma$, define the positive gain of switching player $i$ to the pure strategy $a\in S^i$ by
$$
  A_i(a,\sigma)
  = \max\bigl(0,\bar g^i(\sigma[i\leftarrow\delta_a]) - \bar g^i(\sigma)\bigr).
$$
The Nash map increases the weight of profitable pure deviations and then renormalizes within each player block:
$$
  F(\sigma)^i(a)
  =
  \frac{\sigma^i(a)+A_i(a,\sigma)}
       {\sum_{b\in S^i}\bigl(\sigma^i(b)+A_i(b,\sigma)\bigr)}.
$$
The denominator is at least one, because the original mixed strategy has total mass one and all gain terms are nonnegative.  Hence every block of $F(\sigma)$ is again a point of the corresponding simplex.

The continuity proof follows the syntax of the definition.  Each coordinate is built from coordinate projections, finite sums and products, subtraction, \texttt{max}, and division by a nonzero denominator; Lean packages this in \texttt{nash\_map\_cont}.\href{https://github.com/Solo-ary/Game-Theory-Formalization/blob/camera-ready-icml2026/Brouwer/Gametheory/Nash.lean#L401-L474}{\extlink}  Applying the product version of Brouwer's theorem to this continuous self-map gives a fixed point $F(\sigma)=\sigma$.\href{https://github.com/Solo-ary/Game-Theory-Formalization/blob/camera-ready-icml2026/Brouwer/Gametheory/Nash.lean#L258-L343}{\extlink}

It remains to relate fixed points of $F$ to equilibrium.  If every pure positive gain $A_i(a,\sigma)$ is zero, then every pure deviation is nonprofitable.  By the linearity lemma for \texttt{mixed\_g}, every mixed deviation is a convex combination of pure deviations, so no mixed deviation is profitable.  Conversely, if some pure gain were positive at a fixed point, the normalization equation would force the denominator in player $i$'s block to be strictly larger than one.  The weighted-sum identity for the current mixed strategy then produces a strategy with positive mass and nonpositive gain, contradicting the fixed-point equation for that coordinate.\href{https://github.com/Solo-ary/Game-Theory-Formalization/blob/camera-ready-icml2026/Brouwer/Gametheory/Nash.lean#L477-L581}{\extlink}

\begin{theorem}[Existence of mixed Nash equilibria]
Every finite strategic-form game has a mixed Nash equilibrium.\href{https://github.com/Solo-ary/Game-Theory-Formalization/blob/camera-ready-icml2026/Brouwer/Gametheory/Nash.lean#L477-L581}{\extlink}
\end{theorem}

\section{BrouwerBench: A Lean-Grounded Pilot Benchmark}
\label{sec:benchmark}

The formalization also yields a small benchmark for probing how language
models read and explain Lean-grounded mathematical developments.  We call this
benchmark \emph{BrouwerBench}.  Its purpose is not to measure end-to-end proof
generation: the models are not asked to produce Lean terms that are compiled
by the kernel.  Instead, the benchmark asks whether a model can read local Lean
context, identify the role of definitions and lemmas, and explain how proof
components fit together across the Scarf--Brouwer--Nash pipeline.  We therefore
treat BrouwerBench v1 as a pilot proof-structure probe derived from this
formalization, not as a broad benchmark for general Lean theorem proving.

\subsection{Benchmark Construction}

BrouwerBench v1 contains 80 question--answer items extracted from the
formalization.  The items cover four parts of the development: Scarf's theorem,
the standard-simplex proof of Brouwer's theorem, the product-of-simplices
extension, and the Nash equilibrium application.  Each item contains a short
task-specific Lean-grounded context, a natural-language question, a reference
answer, evidence pointers to the Lean source, and a 0--2 grading rubric.  The
questions are organized by proof role, including definition roles, construction
roles, proof dependencies, parity arguments, analytic limiting steps, theorem
summaries, and endpoint connections.

During evaluation, each model receives the section-level Lean context together
with the task-specific context and question.  The section-level context
includes the relevant definitions, structures, and abbreviations, such as the
Scarf cell/room/door predicates, the finite-grid constructions used in
Brouwer's theorem, the embedding--projection maps for products of simplices,
and the Nash map definitions.  The answer is then scored manually according to
the item rubric: 0 means that the answer misses the main mathematical point, 1
means that it identifies part of the intended structure but omits an essential
dependency, and 2 means that it explains the relevant Lean-grounded proof role
accurately enough to match the reference answer.  This scoring protocol is
intended to support a qualitative pilot evaluation; the limitations of manual
scoring are discussed in Section~\ref{sec:limitations}.

\subsection{Models and Results}

We evaluated four locally deployed models: \texttt{qwen3:8b} as a small
baseline, \texttt{gemma3:12b}, \texttt{kimina-prover:7b}, and
\texttt{gpt-oss:20b}.  The results in
Table~\ref{tab:brouwerbench-v1-models} use the 80-item BrouwerBench v1 dataset,
with a maximum score of 160 points.  The percentage column reports the fraction
of the maximum rubric score.  Runtime is included only as descriptive
information for these local runs.

\begin{table}[H]
  \caption{BrouwerBench v1 results. Scores use a 0--2 rubric per item.}
  \label{tab:brouwerbench-v1-models}
  \begin{center}
    \begin{scriptsize}
      \begin{tabular}{lrrrr}
        \toprule
        Model & Score & Pct. & Time (s) & Avg. (s) \\
        \midrule
        \texttt{gpt-oss:20b} & 122/160 & 76.2\% & 1448.1 & 18.10 \\
        \texttt{qwen3:8b} & 101/160 & 63.1\% & 428.1 & 5.35 \\
        \texttt{gemma3:12b} & 89/160 & 55.6\% & 822.8 & 10.29 \\
        \texttt{kimina-prover:7b} & 58/160 & 36.2\% & 986.4 & 12.33 \\
        \bottomrule
      \end{tabular}
    \end{scriptsize}
  \end{center}
  \vskip -0.1in
\end{table}

\begin{table}[H]
  \caption{BrouwerBench v1 scores by formalization section.}
  \label{tab:brouwerbench-v1-sections}
  \begin{center}
    \begin{scriptsize}
      \begin{tabular}{lrrrr}
        \toprule
        Model & Scarf & Brouwer & Prod. & Nash \\
        \midrule
        \texttt{gpt-oss:20b} & 40/48 & 25/40 & 29/36 & 28/36 \\
        \texttt{qwen3:8b} & 30/48 & 22/40 & 25/36 & 24/36 \\
        \texttt{gemma3:12b} & 25/48 & 21/40 & 24/36 & 19/36 \\
        \texttt{kimina-prover:7b} & 14/48 & 12/40 & 14/36 & 18/36 \\
        \bottomrule
      \end{tabular}
    \end{scriptsize}
  \end{center}
  \vskip -0.1in
\end{table}

In this pilot setting, \texttt{gpt-oss:20b} obtains the highest score,
with 122 out of 160 points.  It scores especially well on theorem-summary and
proof-dependency questions, where the task is to connect named Lean lemmas to
the surrounding proof strategy.  The smaller \texttt{qwen3:8b} baseline obtains
101 points and is comparatively strong on the product and Nash sections.  This
suggests that, for these 80 items, the score is not determined only by model
size.  The lower score of \texttt{kimina-prover:7b} should not be read as a
general statement about prover-oriented models: BrouwerBench asks for
explanatory proof-structure answers rather than low-level proof search or
kernel-checked proof synthesis.

\subsection{Interpretation}

These results should be interpreted as evidence about Lean-grounded proof
understanding within this development, not as direct evidence about the ability
to synthesize complete formal proofs.  A high-scoring answer typically names
the relevant Lean objects and explains why they are needed: for example, why
\texttt{typed\_colorful\_room\_odd c default} uses an inhabited default type,
why Brouwer's limiting argument first stabilizes a color/index set, why
\texttt{project\_embed\_id} is the retraction identity needed in the product
theorem, or how \texttt{mixed\_g\_linear} moves the Nash proof from pure
deviations to mixed deviations.  These are intermediate reasoning skills that
support autoformalization workflows, but they do not by themselves certify a
Lean proof.

BrouwerBench is therefore complementary to end-to-end theorem-proving
evaluations.  End-to-end tasks test whether a model can produce code accepted
by Lean; BrouwerBench tests whether a model can explain the architecture of an
existing formal proof from local formal context.  For a formalization project
like ours, this distinction is useful: many practical proof-assistant workflows
require models to read existing definitions, identify reusable lemmas, explain
failures, and suggest proof plans before a final proof script is produced.

\section{Discussion}
\label{sec:discuss}

The formalization demonstrates that Scarf's combinatorial theorem can serve as
a concrete bridge between finite discrete arguments and classical fixed-point
applications in Lean.  The main technical value of the development lies in the
intermediate proof objects: dominant sets, rooms and doors, nearly colorful
configurations, grid estimates on the standard simplex, and the
embedding--projection construction for products of simplices.  These objects
make the proof route explicit and allow the final Nash equilibrium theorem to
be traced back to a finite parity argument rather than to an opaque invocation
of a fixed-point principle.

This structure is also useful from an AI-for-mathematics perspective.  Large
formal developments require more than isolated theorem proving: a model must
understand how definitions are intended to be used, which lemmas carry the main
mathematical content, and how local proof components support later arguments.
BrouwerBench is designed to probe this kind of proof-structure understanding
within the present development.  Its results suggest that models differ
substantially in their ability to explain Lean-grounded proof architecture, but
the evaluation should be interpreted only as a pilot study.

\subsection{Limitations}
\label{sec:limitations}

The formalization is specialized to one route from Scarf's theorem to
Brouwer's fixed point theorem and Nash equilibrium existence.  Other proofs of
Brouwer's theorem, such as those based on simplicial approximation, degree
theory, or classical topology, are outside the scope of the present
development.  Similarly, the Nash application covers finite games with mixed
strategies; it does not address more general equilibrium concepts or games
with infinite strategy spaces.

BrouwerBench also has important limitations.  The current version contains 80
items, all extracted from a single formalization pipeline.  It is therefore not
large or diverse enough to support broad conclusions about general mathematical
reasoning, Lean theorem proving, or the relative strength of different model
families.  The questions are designed to test local proof-structure
understanding inside this development, not end-to-end synthesis of
kernel-checked Lean proofs.

The scoring protocol is preliminary.  Each answer is graded manually using a
0--2 rubric, and the current version does not report inter-annotator agreement,
adjudication procedures for disputed samples, or a systematic error analysis.
The reported numbers should therefore be read as pilot evidence about the
usefulness of the task format, not as a statistically robust benchmark
comparison.  Future versions should include a larger and more diverse set of
formal developments, multiple graders, explicit disagreement resolution, and
more detailed error categories.

\subsection{Future Work}

There are several natural directions for extending this development. First, the Nash endpoint can be expanded into a broader Lean library for finite game theory. The present formalization establishes the existence of mixed Nash equilibria by reducing finite games to Brouwer's fixed point theorem on a product of simplices. A next step is to build on the same representation of mixed profiles and expected payoffs to formalize more structured classes of games, such as zero-sum games, extensive-form games, repeated games, and equilibrium refinements. This would turn the current endpoint into reusable infrastructure for verified game theory rather than a single isolated existence theorem.

Second, BrouwerBench can be extended from proof-understanding questions to kernel-checked autoformalization tasks. The current version evaluates whether models can identify definitions, dependencies, and proof roles inside an existing Lean development. A richer benchmark could include statement translation, lemma retrieval, proof-script generation, and proof repair from Lean error messages. More broadly, the Scarf--Brouwer--Nash pipeline suggests a reusable formalization pattern: isolate a finite combinatorial core, derive quantitative approximation estimates, and then transfer the result to analysis using compactness and continuity. Applying this pattern to related Sperner- or Scarf-style arguments may help build larger verified libraries at the interface of topology, optimization, and game theory.


\section*{Acknowledgements}

We thank the anonymous reviewers for their constructive comments.
We are grateful to Jiajun Ma for his guidance and support, and to
Haocheng Wang for helpful discussions and feedback on the Lean
development. This work was partially supported by the Xiamen
University Malaysia Research Fund (Grant No. XMUMRF/2025-C15/IMAT/0036).

\section*{Impact Statement}

This paper presents work whose goal is to advance the formalization of mathematics and the evaluation of AI systems for mathematical reasoning. The Lean development may contribute to more reliable and reproducible machine-checked proofs, while the accompanying benchmark may help evaluate autoformalization systems on structured proof developments involving combinatorics, topology, and game theory. However, the benchmark is narrow in scope and should not be interpreted as a comprehensive measure of mathematical reasoning ability. We do not anticipate direct negative societal impacts beyond the general risks of over-reliance on automated reasoning tools without sufficient proof-assistant or human verification.

\bibliographystyle{icml2026}
\bibliography{Brouwer}

\onecolumn
\appendix

\section{Artifact and Version Information}
\label{app:artifact}

The paper-specific Lean artifact is available at the stable tag
\url{https://github.com/Solo-ary/Game-Theory-Formalization/tree/camera-ready-icml2026}.
This repository contains the Lean files and BrouwerBench materials corresponding
to the camera-ready version of this paper. The upstream maintained version of
the formalization is available at \url{https://github.com/math-xmum/Brouwer};
this repository may continue to evolve after publication.

The camera-ready artifact uses Lean 4.22.0, pinned by
\texttt{Brouwer/lean-toolchain}; exact dependency revisions are recorded in
\texttt{Brouwer/lake-manifest.json}. The line counts reported in
Table~\ref{tab:appendix-file-map} are obtained by running \texttt{wc -l} on the
Lean files in this tagged artifact.

The artifact contains two main parts. First, it contains the checked Lean source
for the Scarf--Brouwer--Nash proof pipeline, including the finite Scarf
combinatorics, the standard-simplex Brouwer proof, the product-simplex
reduction, the Nash endpoint, and supporting simplex infrastructure. Second, it
contains the BrouwerBench v1 materials: the 80-item dataset at
\texttt{benchmarks/data/brouwerbench\_v1.jsonl}, section-level context snippets
under \texttt{benchmarks/context/}, raw outputs, scored JSONL files, and reports
under \texttt{benchmarks/results/}, and manual score files under
\texttt{benchmarks/scores/}.

Complete build, validation, model-running, and score-aggregation instructions
are provided in the top-level README and in
\texttt{benchmarks/README.md}.

\section{Formalization Overview}
\label{app:formalization-overview}
The Lean development is organized as a proof pipeline.  The Scarf file
establishes the finite combinatorial theorem; the Brouwer file instantiates it
on finite simplex grids and passes to a limit; the product file transports the
standard-simplex fixed-point theorem to finite products of simplices; and the
Nash file applies that product theorem to mixed strategy spaces.
Table~\ref{tab:appendix-file-map} gives the file-level organization and line
counts used in this paper.

\begin{table}[H]
\caption{File-level structure of the Lean formalization. Line counts are obtained by \texttt{wc -l} on the submitted Lean files.}
\label{tab:appendix-file-map}
\begin{center}
    \begin{small}
      \begin{tabular}{lrll}
\toprule
File & Lines & Main endpoint & Role in pipeline \\
\midrule
\texttt{Scarf.lean} & 2600 & \texttt{Scarf} & finite parity theorem for colorful rooms \\
\texttt{Brouwer.lean} & 794 & \texttt{Brouwer} & grid approximation and limiting fixed point \\
\texttt{Brouwer\_product.lean} & 671 & \texttt{Brouwer\_Product} & retraction from a big simplex to a product \\
\texttt{Nash.lean} & 584 & \texttt{ExistsNashEq} & fixed point implies mixed Nash equilibrium \\
\texttt{Simplex.lean} & 119 & supporting simplex lemmas & reusable simplex infrastructure \\
\midrule
Total & 4768 & -- & complete Scarf--Brouwer--Nash pipeline \\
\bottomrule
\end{tabular}
    \end{small}
  \end{center}
  \vskip -0.1in
\end{table}

The formalization also separates reusable infrastructure from proof endpoints. For example, the Scarf layer introduces an indexed family of linear orders before specifying any simplex grid, while the product layer proves its embedding and projection lemmas before the Nash layer uses products of simplices as mixed strategy spaces. This separation is mirrored by BrouwerBench: many benchmark items ask why an intermediate object is needed before the endpoint theorem can be applied.

\section{Scarf Proof Components}
\label{app:scarf-details}

Table~\ref{tab:appendix-scarf-components} summarizes the combinatorial objects that carry the Scarf proof. The first layer is order theoretic: \texttt{isDominant} says that a finite set of points is controlled by a finite set of order indices. The second layer refines dominant pairs into cells, rooms, and doors by cardinality. The third layer equips these pairs with a coloring and counts typed nearly colorful incidences.

\begin{table}[H]
\caption{Scarf proof components and their roles.}
\label{tab:appendix-scarf-components}
\begin{center}
    \begin{small}
      \begin{tabular}{lll}
\toprule
Lean object & Proof role & Used to establish \\
\midrule
\texttt{IndexedLOrder} & family of orders indexed by colors & dominance relation \\
\texttt{isDominant} & order-theoretic cell condition & finite cell structure \\
\texttt{isCell}, \texttt{isRoom}, \texttt{isDoor} & cardinality refinements & room--door incidence graph \\
\texttt{isDoorof} & relation between a door and adjacent room & incidence counting \\
\texttt{isOutsideDoor}, \texttt{isInternalDoor} & boundary/interior split & degree one versus degree two \\
\texttt{isColorful} & cell has exactly its color image & target objects \\
\texttt{isNearlyColorful}, \texttt{isTypedNC} & one fixed missing color & fiberwise parity argument \\
\texttt{dbcountingset} & typed door--room incidences & double counting \\
\texttt{typed\_colorful\_room\_odd} & odd colorful-room count & nonempty colorful filter \\
\texttt{Scarf} & endpoint theorem & colorful room exists \\
\bottomrule
\end{tabular}
    \end{small}
  \end{center}
  \vskip -0.1in
\end{table}

The room--door degree facts are the local combinatorial core. Outside doors have one adjacent room, while internal doors have two. After fixing a missing type, the parity proof partitions \texttt{dbcountingset} on the door side into outside and internal doors, and on the room side into colorful and non-colorful typed nearly colorful rooms. The outside contribution is odd, while the internal-door and non-colorful-room contributions are even. The remaining colorful-room contribution is therefore odd and hence nonzero.

\section{Brouwer Proof Components}
\label{app:brouwer-details}

The standard-simplex proof converts the finite Scarf theorem into an analytic fixed point. Table~\ref{tab:appendix-brouwer-components} records the main proof dependencies.

\begin{table}[H]
\caption{Key objects in the Scarf-to-Brouwer argument.}
\label{tab:appendix-brouwer-components}
\begin{center}
    \begin{small}
      \begin{tabular}{lll}
\toprule
Lean object & Local meaning & Global use \\
\midrule
\texttt{TT n l} & finite integer grid with coordinate sum \texttt{l} & finite domain for Scarf \\
\texttt{TTtostdSimplex} & divides grid coordinates by \texttt{l} & realizes grid points in the simplex \\
\texttt{TT.ILO} & indexed lexicographic orders on the grid & Scarf order structure \\
\texttt{Fcolor} & coordinate with \texttt{x\_i <= f(x)\_i} & grid coloring \\
\texttt{room\_seq} & colorful room at each grid level & discrete approximations \\
\texttt{room\_point\_seq} & selected point from each room & sequence to compactify \\
\texttt{gpkg} & fixed color/index set and subsequence & stable coordinate split \\
\texttt{hpkg} & convergent subsequence package & limit point \texttt{z} \\
\texttt{tendsto\_diam\_to\_zero} & rooms shrink along the grid & transfer between room points \\
\texttt{dominant\_coords\_tend\_to\_zero} & coordinates outside fixed set vanish & outside-support control \\
\texttt{f\_coords\_ge\_z\_coords} & inequalities pass to the limit & endpoint equality argument \\
\texttt{Brouwer} & standard-simplex theorem & fixed point \texttt{f z = z} \\
\bottomrule
\end{tabular}
    \end{small}
  \end{center}
  \vskip -0.1in
\end{table}

The final proof endpoint uses two kinds of estimates. On the stabilized color/index set \(C\), colorful witnesses and continuity give coordinate inequalities \((f z)_i \geq z_i\). Outside \(C\), dominance estimates show that the coordinates of \(z\) vanish. Since \(z\) and \(f z\) both lie in the simplex, the sum-one constraints leave no extra mass: the inequalities on \(C\) become equalities, the outside coordinates are zero, and extensionality yields \(f z = z\).

\section{Product-Simplex Proof Components}
\label{app:product-details}

The product extension reduces a fixed-point problem on a finite product of simplices to a fixed-point problem on one larger simplex. The reduction requires explicit coordinate bookkeeping and a projection whose blockwise denominators are positive.

\begin{table}[H]
\caption{Product-simplex constructions.}
\label{tab:appendix-product-components}
\begin{center}
    \begin{small}
      \begin{tabular}{lll}
\toprule
Lean object & Purpose & Key dependency \\
\midrule
\texttt{index\_split}, \texttt{index\_combine} &
convert flat and blockwise indices &
inverse lemmas for sums \\
\texttt{blockSum}, \texttt{blockWeight} &
measure mass in each block &
positive target block weights \\
\texttt{deficit}, \texttt{tPush} &
measure and bound mass correction &
continuous push amount \\
\texttt{pushTowardsZ} &
move toward a uniform interior point &
positive post-push block sums \\
\texttt{project\_to\_product} &
normalize each block &
nonzero block denominators \\
\texttt{embed\_from\_product} &
place product coordinates in big simplex &
prescribed block masses \\
\texttt{project\_embed\_id} &
projection after embedding is identity &
fixed-point transport \\
\texttt{Brouwer\_Product} &
product fixed-point theorem &
lift, apply Brouwer, project back \\
\bottomrule
\end{tabular}
    \end{small}
  \end{center}
  \vskip -0.1in
\end{table}

The positivity argument is the nontrivial part of projection. A raw point in the big simplex may have insufficient mass in a block, so blockwise normalization cannot be treated as ordinary coordinate projection. The push construction moves points toward a uniform interior point. If the push amount is positive, positive block weight contributes directly. If the push amount is zero, zero deficit shows that the original block already meets its positive target weight. These cases justify the denominator in \texttt{project\_to\_product} and allow the continuity proof to use division by a nonzero continuous function.

\section{Nash Proof Components}
\label{app:nash-details}

The Nash layer turns the product fixed-point theorem into equilibrium existence. The finite game supplies one simplex per player, and the Nash map raises weights on profitable pure deviations before renormalizing each player block.

\begin{table}[H]
\caption{Objects in the Nash endpoint.}
\label{tab:appendix-nash-components}
\begin{center}
    \begin{small}
      \begin{tabular}{lll}
\toprule
Lean object & Local meaning & Endpoint role \\
\midrule
\texttt{Game}, \texttt{FinGame} & strategic-form data with finiteness & finite mixed space \\
\texttt{mixedS} & product of player simplices & fixed-point domain \\
\texttt{mixed\_g} & expected payoff & payoff comparison \\
\texttt{g\_function} & current weight plus positive payoff gain & Nash-map numerator \\
\texttt{nash\_map\_aux} & normalized coordinate & simplex coordinate \\
\texttt{nash\_map\_cert} & sum-one and nonnegativity certificate & mixed profile output \\
\texttt{nash\_map\_cont} & continuity of Nash map & Brouwer hypothesis \\
\texttt{mixed\_g\_linear} & mixed payoff as pure-deviation average & pure-to-mixed step \\
\texttt{self\_div\_lemma} & normalization contradiction & rules out profitable pure deviations \\
\texttt{ExistsNashEq} & endpoint theorem & mixed Nash equilibrium exists \\
\bottomrule
\end{tabular}
    \end{small}
  \end{center}
  \vskip -0.1in
\end{table}

At a fixed point of \texttt{nash\_map}, a profitable pure deviation would make the normalization denominator strictly larger than one. The proof then finds a nonzero coordinate with no positive improvement. Fixedness gives an equality of the form \(\sigma_s=\sigma_s/\mathrm{sum\_g}\), and \texttt{self\_div\_lemma} forces \(\mathrm{sum\_g}=1\), contradicting the strict inequality. The remaining pure-deviation bound extends to arbitrary mixed deviations through \texttt{mixed\_g\_linear}.

\clearpage
\section{Selected Lean Interfaces}
\label{app:selected-interfaces}

This appendix records selected interfaces rather than full proof scripts. The complete proof terms and auxiliary lemmas are available in the repository; the snippets below show the formal boundaries between the main layers of the development.

\subsection{Scarf's Combinatorial Core}

\begin{lstlisting}[language=lean,basicstyle=\scriptsize\ttfamily,caption={Representative interfaces for indexed orders, dominance, and Scarf's theorem.},label={lst:app-scarf-interface}]
class IndexedLOrder (I T : Type*) where
  IST : I → LinearOrder T
def isDominant (σ : Finset T) (C : Finset I) : Prop :=
  ∀ y, ∃ i ∈ C, ∀ x ∈ σ, y ≤[i] x
abbrev isRoom (σ : Finset T) (C : Finset I) : Prop :=
  isCell σ C ∧ C.card = σ.card
abbrev isDoor (σ : Finset T) (C : Finset I) : Prop :=
  isCell σ C ∧ C.card = σ.card + 1
theorem internal_door_two_rooms [Fintype T]
    (τ : Finset T) (D : Finset I)
    (h : IST.isInternalDoor τ D) :
    ∃ (σ₁ σ₂ : Finset T) (C₁ C₂ : Finset I),
      (σ₁, C₁) ≠ (σ₂, C₂) ∧
      IST.isRoom σ₁ C₁ ∧ IST.isRoom σ₂ C₂ ∧
      isDoorof τ D σ₁ C₁ ∧ isDoorof τ D σ₂ C₂ ∧
      (∀ σ C, IST.isRoom σ C → isDoorof τ D σ C →
        (σ = σ₁ ∧ C = C₁) ∨ (σ = σ₂ ∧ C = C₂)) := by
  ...
lemma NC_or_C_of_door
    (h1 : isTypedNC c i τ D) (h2 : isDoorof τ D σ C) :
    isTypedNC c i σ C ∨ isColorful c σ C := by
  ...
lemma typed_colorful_room_odd (i : I) :
    Odd (Finset.filter
      (fun x => isColorful c x.2.1 x.2.2)
      (dbcountingset c i)).card := by
  ...
theorem Scarf : (IST.colorful c).Nonempty := by
  ...
\end{lstlisting}

\newpage
\subsection{Brouwer from Scarf}

\begin{lstlisting}[language=lean,basicstyle=\scriptsize\ttfamily,caption={Representative interfaces for the grid construction and Brouwer endpoint.},label={lst:app-brouwer-interface}]
abbrev TT :=
  {x : Πₗ (_ : Fin n), Fin (l+1) |
    ∑ i, (x i : ℕ) = l}

def TTtostdSimplex (x : TT n l) : stdSimplex ℝ (Fin n) :=
  ...

def Fcolor (x : TT n l) : Fin n :=
  stdSimplex.pick x (f x)

theorem size_bound_in
    (σ : Finset (TT n l)) (C : Finset (Fin n))
    (h : TT.ILO.isDominant σ C) :
    ∀ x ∈ σ, ∀ y ∈ σ, ∀ i,
      |(x : stdSimplex ℝ (Fin n)) i -
       (y : stdSimplex ℝ (Fin n)) i| < 2 * (n + 1) / l := by
  ...

theorem size_bound_out
    (σ : Finset (TT n l)) (C : Finset (Fin n))
    (h : TT.ILO.isDominant σ C) :
    ∀ x ∈ σ, ∀ i ∉ C,
      (x : stdSimplex ℝ (Fin n)) i < (n + 1) / l := by
  ...

theorem f_coords_ge_z_coords
    (f : stdSimplex ℝ (Fin n) → stdSimplex ℝ (Fin n))
    (hf : Continuous f) :
    ∀ i ∈ (gpkg f).1.1,
      (f (hpkg f).1.1).1 i ≥ ((hpkg f).1.1).1 i := by
  ...

theorem Brouwer
    (f : stdSimplex ℝ (Fin n) → stdSimplex ℝ (Fin n))
    (hf : Continuous f) : ∃ x, f x = x := by
  ...
\end{lstlisting}

\subsection{Product Simplices}

\begin{lstlisting}[language=lean,basicstyle=\scriptsize\ttfamily,caption={Representative interfaces for the embedding--projection reduction.},label={lst:app-product-interface}]
abbrev BigSimplex := stdSimplex ℝ (Fin (total_card card))

abbrev ProductSimplices :=
  (i : I) → stdSimplex ℝ (Fin (card i))

lemma project_embed_id (y : ProductSimplices card) :
    project_to_product card (embed_from_product card y) = y := by
  ...

lemma project_continuous :
    Continuous (project_to_product card) := by
  ...

lemma embed_continuous :
    Continuous (embed_from_product card) := by
  ...

theorem Brouwer_Product
    (f : ProductSimplices card → ProductSimplices card)
    (hf : Continuous f) :
    ∃ x : ProductSimplices card, f x = x := by
  ...
\end{lstlisting}

\newpage
\subsection{Nash Equilibria}
\label{app:nash-interface}

\begin{lstlisting}[language=lean,basicstyle=\scriptsize\ttfamily,caption={Representative interfaces for the Nash endpoint.},label={lst:app-nash-interface}]
variable (α : Type*) [Fintype α] [DecidableEq α]
abbrev S:= stdSimplex ℝ α
structure Game where
    I : Type*           -- The set of player
    --deEqI : DecidableEq I := inferInstance -- Decidable Eq
    HI : Inhabited I     -- at least one player
    SS : I → Type*       -- S is the set of strategies
    HSS (i :I) : Inhabited (SS i) -- The set of strategies is nonempty
    --deEqSS (i : I) : DecidableEq (SS i)
    g : I → (Π i, SS i) →  ℝ
    -- an elements in Π i, SS is a move of all players.
    -- g i is the payoff of the i-th player
structure FinGame extends Game where
  FinI : Fintype I
  FinSS : ∀ i : I , Fintype (SS i)

namespace FinGame
variable {G : FinGame}
variable (G) in
abbrev mixedS  := (i : G.I) → stdSimplex ℝ (G.SS i)

def mixed_g (i : G.I) (m : Π i, S (G.SS i) ) : ℝ := ∑ s : (Π j, G.SS j) , (∏ j,  m j (s j)) * (G.g i s)

lemma mixed_g_linear : G.mixed_g i (update  x i y) = ∑ s : G.SS i, y s * G.mixed_g i (update x i (stdSimplex.pure s)) := by
  ...

def mixedNashEquilibrium {G: FinGame} (x : G.mixedS) :=
  ∀ (i:G.I), ∀ (y : S (G.SS  i)),
     G.mixed_g i x ≥ G.mixed_g i (update  x i y)

end FinGame
section mixedNashEquilibrium
variable (G : FinGame)
open FinGame
variable {G}

noncomputable abbrev g_function (i : G.I) (σ : G.mixedS) (a : G.SS i) : ℝ :=
  σ i a + max 0 (mixed_g i (Function.update σ i (stdSimplex.pure a)) - mixed_g i σ)

lemma one_le_sum_g (i : G.I) (σ : G.mixedS) :
    1 ≤ ∑ b : G.SS i, g_function i σ b := by
  ...

noncomputable abbrev nash_map_aux (σ : G.mixedS) (i : G.I) (a : G.SS i) : ℝ :=
  g_function i σ a / ∑ b : G.SS i, g_function i σ b

lemma nash_map_cert (σ : G.mixedS) (i : G.I) :
  (nash_map_aux σ i) ∈ S (G.SS i) := by
  ...

variable (G)

noncomputable def nash_map (σ: G.mixedS) : G.mixedS :=
  fun (i : G.I) ↦ ⟨nash_map_aux σ i, nash_map_cert σ i⟩

lemma nash_map_cont : Continuous $ nash_map G :=
  by
  ...

theorem ExistsNashEq : ∃ σ : G.mixedS , mixedNashEquilibrium σ := by {
  ...
}

end mixedNashEquilibrium
\end{lstlisting}

\section{BrouwerBench Dataset}
\label{app:dataset}

BrouwerBench v1 is stored as JSONL. Every row corresponds to one hand-checkable question about a local proof role in the formalization. The 80 items are distributed across the same four proof layers as the paper.

\begin{table}[H]
\caption{BrouwerBench v1 coverage by formalization section.}
\label{tab:appendix-dataset-coverage}
\begin{center}
    \begin{small}
      \begin{tabular}{lrrrrr}
\toprule
Section & Scarf & Brouwer & Product Brouwer & Nash & Total \\
\midrule
Items & 24 & 20 & 18 & 18 & 80 \\
\bottomrule
\end{tabular}
    \end{small}
  \end{center}
  \vskip -0.1in
\end{table}

\begin{table}[H]
\caption{BrouwerBench v1 task types.}
\label{tab:appendix-task-types}
\begin{center}
    \begin{small}
      \begin{tabular}{lrrrrrrr}
\toprule
Type & Def. role & Proof dep. & Parity & Summary & Constr. role & Analysis & Endpoint \\
\midrule
Items & 7 & 28 & 4 & 3 & 17 & 11 & 10 \\
\bottomrule
\end{tabular}
    \end{small}
  \end{center}
  \vskip -0.1in
\end{table}

\subsection{Dataset Schema}

Each item contains the fields listed in Table~\ref{tab:appendix-dataset-schema}. The \texttt{context} and \texttt{question} fields are shown to the model together with a section-level Lean prelude. The reference answer and rubric are used for grading rather than included in the prompt.

\begin{table}[H]
\caption{Fields in one BrouwerBench JSONL item.}
\label{tab:appendix-dataset-schema}
\begin{center}
    \begin{small}
      \begin{tabular}{ll}
\toprule
Field & Meaning \\
\midrule
\texttt{id} & stable item identifier \\
\texttt{section} & one of \texttt{Scarf}, \texttt{Brouwer}, \texttt{Brouwer\_product}, or \texttt{Nash} \\
\texttt{task\_type} & proof-role category used for analysis \\
\texttt{source} & Lean source file from which the item is drawn \\
\texttt{section\_prelude} & Lean-style declaration snippet prepended to the prompt \\
\texttt{context} & compact task-specific formal context \\
\texttt{question} & natural-language query answered by the model \\
\texttt{gold\_answer} & reference explanation used during grading \\
\texttt{evidence} & file-line anchors used to create and check the item \\
\texttt{rubric} & item-specific 0--2 criteria and required ideas \\
\bottomrule
\end{tabular}
    \end{small}
  \end{center}
  \vskip -0.1in
\end{table}

\subsection{Example Items}

\paragraph{Scarf definition-role item.}
Item \texttt{scarf\_001} supplies the local context
\begin{quote}
\texttt{isCell} abbreviates \texttt{isDominant}; \texttt{isRoom} adds equal point/index cardinalities; \texttt{isDoor} adds one extra index.
\end{quote}
It asks how \texttt{isDominant}, \texttt{isCell}, \texttt{isRoom}, and \texttt{isDoor} relate. A score-2 answer must explain dominance as the core cell condition, state the two cardinality refinements, and connect them to the room--door parity structure.

\paragraph{Product endpoint item.}
Item \texttt{product\_011} provides the lifted map
\[
\begin{aligned}
\texttt{f\_lifted}
&=
\texttt{embed\_from\_product} \\
&\quad \circ f \circ \texttt{project\_to\_product}
\end{aligned}
\]
and asks why the fixed point is first found in \texttt{BigSimplex} rather than directly in \texttt{ProductSimplices}. A score-2 answer must explain that the available Brouwer theorem is applied to the lifted single-simplex map, then projection and \texttt{project\_embed\_id} transport the equation back to the product.

\paragraph{Nash endpoint item.}
Item \texttt{nash\_011} gives the contradiction branch in \texttt{ExistsNashEq}: the hypothesis \(1<\mathrm{sum\_g}\), the fixed-point equation for \texttt{nash\_map}, and \texttt{self\_div\_lemma}. A complete answer explains how a nonzero coordinate gives \(\sigma_s=\sigma_s/\mathrm{sum\_g}\), why \texttt{self\_div\_lemma} forces \(\mathrm{sum\_g}=1\), and how this contradicts the profitable-deviation branch.

\section{Scoring Protocol}
\label{app:scoring}

The benchmark uses manual item-level scoring. Each item has a rubric tailored to its proof role, but all rubrics follow the shared scale in Table~\ref{tab:appendix-score-scale}.

\begin{table}[H]
\caption{Shared BrouwerBench score scale.}
\label{tab:appendix-score-scale}
\begin{center}
    \begin{small}
      \begin{tabular}{lll}
\toprule
Score & Interpretation & Typical case \\
\midrule
0 & misses the formal proof role & generic Brouwer or Nash explanation unrelated to the given Lean context \\
1 & partially identifies the role & names a subsequence but misses why the fixed set \(C\) is needed \\
2 & explains the Lean-grounded role & connects the named object, dependency, and proof endpoint \\
\bottomrule
\end{tabular}
    \end{small}
  \end{center}
  \vskip -0.1in
\end{table}

The grading distinguishes a mathematically plausible answer from a formalization-specific one. An answer that correctly describes a classical theorem but ignores the supplied Lean objects receives partial credit only when it still captures an essential proof dependency. Hallucinated Lean claims, such as applying the wrong endpoint theorem or inventing an unsupported proof route, lower the score. Exact spelling of every Lean identifier is not required when the mathematical dependency is unambiguous, but named objects matter when the question is specifically about their proof role.

The current study uses one manual score file per model output. This makes errors auditable at item level: the scored JSONL artifacts store the answer, rubric, score, and score note together. As discussed in \S\ref{sec:discuss}, a larger benchmark should add independent graders and adjudication for borderline partial-credit cases.

\section{Detailed Benchmark Results}
\label{app:results}

The main text reports overall and section-level scores. Table~\ref{tab:appendix-results-by-type} gives the task-type breakdown. It shows that the evaluation probes several kinds of proof understanding rather than a single style of question.

\begin{table}[H]
\caption{BrouwerBench v1 scores by task type.}
\label{tab:appendix-results-by-type}
\begin{center}
    \begin{small}
      \begin{tabular}{lrrrrrrr}
\toprule
Model & Def. & Dep. & Parity & Summary & Constr. & Analysis & Endpoint \\
\midrule
\texttt{gpt-oss:20b} & 9/14 & 45/56 & 7/8 & 6/6 & 26/34 & 15/22 & 14/20 \\
\texttt{qwen3:8b} & 8/14 & 38/56 & 5/8 & 4/6 & 23/34 & 13/22 & 10/20 \\
\texttt{gemma3:12b} & 5/14 & 34/56 & 5/8 & 4/6 & 20/34 & 11/22 & 10/20 \\
\texttt{kimina-prover:7b} & 6/14 & 20/56 & 1/8 & 1/6 & 16/34 & 9/22 & 5/20 \\
\bottomrule
\end{tabular}
    \end{small}
  \end{center}
  \vskip -0.1in
\end{table}

\begin{table}[H]
\caption{Reported model artifact stems and decoding settings.}
\label{tab:appendix-model-artifacts}
\begin{center}
    \begin{small}
      \begin{tabular}{lp{0.28\textwidth}p{0.30\textwidth}}
\toprule
Model & Shared artifact stem & Decoding note \\
\midrule
\texttt{gpt-oss:20b} &
\texttt{gpt-oss\_20b\_np4096} &
\texttt{num\_predict=4096} \\
\texttt{qwen3:8b} &
\texttt{qwen3\_8b} &
default short-answer run \\
\texttt{gemma3:12b} &
\texttt{gemma3\_12b} &
default short-answer run \\
\texttt{kimina-prover:7b} &
\texttt{kimina-prover\_7b} &
default short-answer run \\
\bottomrule
\end{tabular}
    \end{small}
  \end{center}
  \vskip -0.1in
\end{table}

The stems in Table~\ref{tab:appendix-model-artifacts} appear after the shared dataset prefix \texttt{brouwerbench\_v1\_\_}. Raw outputs use the result directory and the \texttt{.jsonl} suffix, while manual score files use the score directory and the \texttt{.manual.jsonl} suffix. The full per-item responses, manual score notes, scored JSONL files, and generated Markdown reports are retained in the benchmark artifact directory. This keeps the paper tables compact while preserving the item-level audit trail used for the reported totals.

\newpage
\section{Reproducibility Notes}
\label{app:reproducibility}

The reported runs were produced locally with Ollama on an Apple M4 Pro machine with 24 GB memory. Lean 4.22.0 is pinned by \texttt{Brouwer/lean-toolchain}, and package revisions are recorded in \texttt{Brouwer/lake-manifest.json}. The benchmark artifact records the dataset at \texttt{benchmarks/data/brouwerbench\_v1.jsonl}, raw outputs and generated reports under \texttt{benchmarks/results/}, and manual score files under \texttt{benchmarks/scores/}.

Benchmark validation checks JSONL parsing, unique ids, rubric fields, source and prelude links, evidence anchors, and whether task contexts leak proof-body material. Detailed command-line instructions for reproducing validation, local model runs, and score aggregation are provided in the repository README.

The reported GPT run uses a single longer output budget, \texttt{num\_predict=4096}, for all 80 items. Row \texttt{product\_008} still exhausts the thinking budget, so its empty response is retained and scored as a failure.

Section-level prompt preludes are stored as Lean-style snippets rather than standalone Lean modules. They expose the relevant definitions, structures, abbreviations, and inductive declarations to the model while avoiding proof-body leakage. The snippets are therefore prompt context artifacts; they are validated by the benchmark tooling but are not compiled as independent Lean files.
\end{document}